\documentclass[12pt]{article}
\usepackage{graphics}
\renewcommand{\theequation}{\arabic{section}.\arabic{equation}}
\begin{document}
%------------------------------------------------------------------------------
\markright{Instability of ...}
%------------------------------------------------------------------------------
\title{Instability of Extremal Relativistic Charged Spheres}

\author{Peter Anninos~$^*$ and Tony Rothman $^\dagger$
\\[2mm]
{\small\it \thanks{anninos1@llnl.gov}}~ \it University of
California, \\ \it Lawrence Livermore National Laboratory,
    Livermore CA 94550
\\
{\small\it\thanks{trothman@brynmawr.edu}}~ \it Department of Physics,
\it Bryn Mawr College, \\
\it 101 N. Merion Ave., 
\it Bryn Mawr, PA 19010 }

\date{{\small   \LaTeX-ed \today}}
%-----------------------------------------------------------------------------

\maketitle
%-----------------------------------------------------------------------------

%-----------------------------------------------------------------------------
\begin{abstract}
With the question, ``Can relativistic charged spheres form
extremal black holes?" in mind, we investigate the properties of
such spheres from a classical point of view.  The investigation is
carried out numerically by integrating the
Oppenheimer-Volkov equation for relativistic charged fluid spheres
and finding interior Reissner-Nordstr\"om solutions for these
objects. We consider both constant density and adiabatic equations
of state, as well as several possible charge distributions, and
examine stability by both a normal mode and an energy analysis. In
all cases, the stability limit for these spheres lies between the
extremal ($Q = M$) limit and the black hole limit ($R = R_+$).
That is, we find that charged spheres undergo gravitational
collapse before they reach $Q = M$, suggesting that extremal
Reissner-Nordtr\"om black holes produced by collapse are ruled
out. A general proof of this statement would support a strong form
of the cosmic censorship hypothesis, excluding not only stable
naked singularities, but stable extremal black holes. The
numerical results also indicate that although the interior
mass-energy $m(R)$ obeys the usual $m/R <  4/9$ stability limit
for the Schwarzschild interior solution, the gravitational mass
$M$ does not.  Indeed, the stability limit approaches $R_+$ as $Q
\to M$. In the Appendix we also argue that Hawking radiation will
not lead to an extremal Reissner-Nordstr\"om black hole.  All our
results are consistent with the third law of black hole dynamics,
as currently understood.

 \vspace*{5mm} \noindent PACS: 04.70, 04.70 Bw,
97.60.Lf,
\\ Keywords: Extremal Black holes, Reissner-Nordstr\" om solution,
Stellar Stability, Oppenheimer-Volkov Equation.
\end{abstract}

%-----------------------------------------------------------------------------
\section{Introduction}
\label{sec1}
%-----------------------------------------------------------------------------
Extremal black holes, black holes for which the charge or angular
momentum parameter equals the mass, occupy an exceptional position
in black hole thermodynamics: due to their vanishing surface
gravity they represent the absolute zero state of black hole
physics. Over the past five or six years, compelling evidence has
accumulated that extremal black holes represent a fundamentally
different class of objects than their nonextremal counterparts. In
some respects this is not surprising because the horizon structure
changes completely at extremality, (see, eg. \cite{Hawking73}) and
all the thermodynamic properties of black holes depend on the
horizon structure.  The exact nature of extremality is still the
subject of some debate. Semi-classical calculations for eternal,
extremal black holes \cite{HHR95,Teit95} indicate that the entropy
is zero, while more recent semi-classical calculations for
extremal objects collapsing into black holes
\cite{LRS00,Rothman00} find that the temperature, and hence
entropy, is not well defined. Similar results have been found for
BTZ black holes \cite{Medved01}. String theory calculations for
extremal black holes, on the other hand \cite{Strom96}, have
yielded the ordinary Bekenstein-Hawking value for the entropy.

Given the pathological properties of extremal black holes, at
least from the semi-classical side, one reasonable supposition is
that such objects are, for reasons as yet unclear, disallowed by
nature. Israel \cite{Israel86}, for example, has proven a form of
the third law, which shows that it is impossible to reach
extremality by any finite-time, continuous accretion process (with
some caveats). On the other hand, some simple solutions of the
Einstein equations for extremally charged objects are known
\cite{Bou73,FH79,P83}. These studies, however, consider the rather
unrealistic scenario of collapsing, infinitely thin shells.
Moreover, the extremal solutions are evidently unstable, which
again casts doubt on their relevance as models for genuine
physical objects.

In this paper, we take a purely classical point of view and
consider the stability of charged, spherical matter distributions
that satisfy the Einstein equations. Essentially we investigate
interior Reissner-Nordstr\"om solutions, focusing on the extremal
$Q = M$ case. A study of relativistic charged spheres was
undertaken thirty years ago by Bekenstein \cite{Beck71}, but
without numerics he was unable to reach many firm conclusions
about the ability of charge to prohibit collapse. More recently a
numerical investigation has been carried out by de Felice et al.
\cite{deF99} (henceforth dFSY). The dFSY study was limited in that
it considered only spheres with constant matter density $\rho$ and
a power-law charge distribution. Furthermore, although $\rho$ was
taken to be constant in deriving the equation of hydrostatic
equilibrium, the stability analysis assumed an adiabatic equation
of state, i.e., $p \propto \rho_{rm}^\gamma$, where $\rho_{rm}$ is
the rest mass density. As will become clear below, the $\rho =
\mbox{constant}$ case is the only convenient one for numerical
integration; nevertheless, it is possible to construct a scheme
capable of handling the more general adiabatic equation of state,
as well as other charge distributions.  We have done this for the
present study and find that in all cases examined, instability
sets in before extremality is reached.

%-----------------------------------------------------------------------------
\section{Einstein and Oppenheimer-Volkov equations for charged spheres}
\setcounter{equation}{0} \label{sec2}
%-----------------------------------------------------------------------------

To derive the relativistic hydro- and electro-dynamic equations
for a charged fluid sphere, we assume a general spherically
symmetric metric in the form
\begin{equation}
ds^2 = -e^{2\Phi}{\rm d}t^2 + e^{2\Lambda}dr^2 + r^2{\rm
d}\theta^2 + r^2{\rm sin^2}\theta d\phi^2, \label{metric}
\end{equation}
where $\Phi = \Phi_0(r) + \Phi_1(r,t)$ and $\Lambda = \Lambda_0(r)
+ \Lambda_1(r,t)$. In this section we are concerned only with the
zeroeth-order (static) Einstein equations and so set $\Phi_1 =
\Lambda_1 = 0$.  The first-order quantities will be employed for
the stability analysis in \S \ref{sec4}. We write the Einstein
equations in the form
\begin{equation}
G_{\mu\nu} \equiv R_{\mu\nu} - \frac{1}{2}g_{\mu\nu}R
           = 8\pi T_{\mu\nu},\label{einstein}
\end{equation}
with $c=G=1$. For charged spheres, the energy-momentum tensor will
consist of a perfect-fluid part
\begin{equation}
(T_{\mu\nu})_{hydro} = (\rho + p)u_\mu u_\nu +
pg_{\mu\nu}\label{Thyd}
\end{equation}
plus an electromagnetic part
\begin{equation}
4\pi (T_{\mu\nu})_{EM} = F_\mu^{\;\alpha} F_{\nu\alpha} -
\frac{1}{4}g_{\mu\nu}F_{\alpha\beta}F^{\alpha\beta}. \label{TEM}
\end{equation}
In these expressions $\rho = \rho_{rm} + e$ is the total mass
density, $\rho_{rm}$ is the rest mass density, $e$ is the internal
energy density, $p$ is the fluid pressure, $u_\mu$ is the
four-velocity, and $F_{\mu\nu}$ is the electromagnetic field
strength tensor. As in the derivation of the exterior
Reissner-Nordstr\"om solution, we take the magnetic field $B$ to
be zero.  Due to spherical symmetry, the electric field $E$ must
be entirely radial and the electromagnetic field tensor
$F_{\mu\nu}$, which has only two components, $F_{01} = -E$ and
$F_{10} = E$, must satisfy the Maxwell equations
\begin{equation}
\frac{1}{\sqrt{-g}}(\sqrt{-g}F^{\mu\nu}),_\nu = 4\pi j^\mu,
\label{maxwell}
\end{equation}
where $j^\mu$ is the four-current. The only nonvanishing
derivative for the static case is $\nu = r$. Working out
(\ref{maxwell}) gives
\begin{equation}
E(r) = \frac{e^{\Phi(r) + \Lambda(r)}Q(r)}{r^2}, \label{E}
\end{equation}
where
\begin{equation}
Q(r)\equiv 4\pi \int_0^r e^{\Phi(r') + \Lambda(r')}r'^2 j^0(r')dr'
\label{Q}.
\end{equation}
Eq. (\ref{E}) immediately implies
 \begin{equation}
F^{01} = \frac{e^{-(\Phi(r) + \Lambda(r))}Q(r)}{r^2}. \label{F}
\end{equation}
Note that (\ref{E}) and (\ref{F}) are  of exactly the same form as
the exterior Reissner-Nordstr\"om solution, as they must be by
Gauss' law.  The only difference is that for the interior case $Q$
represents $Q(r)$ rather than the total charge.
$(T_{\mu\nu})_{EM}$ must also be of the standard
Reissner-Nordstr\"om form, or from (\ref{TEM}) and (\ref{F})
\begin{equation}
(T_{\mu\nu})_{EM} = \frac{Q^2(r)}{8\pi r^4}\;\rm {diag}
[~e^{2\Phi},\; -e^{2\Lambda},\; r^2,\;
r^2\rm{sin}^2\theta~].\label{TEM2}
\end{equation}
Also note that $T_{EM} \equiv (T^\mu_{\ \mu})_{EM} = 0$ due to the
antisymmetry of $F_{\mu\nu}$.

The left-hand-side of the Einstein equations are the same as for
any static, spherically symmetric solution. With the above
expressions for $T_{\mu\nu}$, the $(00)$ equation is found to be
\begin{equation}
\frac{1}{r^2} + \frac{2\Lambda' e^{-2\Lambda}}{r} -
\frac{e^{-2\Lambda}}{r^2}
     = 8\pi \rho + \frac{Q^2}{r^4},
     \label{00}
\end{equation}
and the $(11)$ equation is
\begin{equation}
\frac{1}{r^2} - \frac{2\Phi' e^{-2\Lambda}}{r} -
\frac{e^{-2\Lambda}}{r^2}
     = \frac{Q^2}{r^4} - 8\pi p,
     \label{11}
\end{equation}
where (${}'$) denotes derivative with respect to $r$.

If one multiplies (\ref{00}) by $r^2$, one gets
\[
\frac{d (e^{-2\Lambda} r)}{dr} = 1 - 8\pi\rho r^2 -
\frac{Q^2(r)}{r^2},
\]
or,
\begin{equation}
e^{-2\Lambda} = 1 - \frac{2 m(r)}{r}-
 \frac{{\cal F}(r)}{r},\label{eL}
\end{equation}
where
\[
 m(r) \equiv 4\pi\int_0^r\rho\; r'^2\; dr' \ \  \ \ \mbox{and}\ \
 \ \ {\cal F}(r) \equiv \int_0^r\frac{Q^2(r')}{r'^2}\; dr'.
 \]
In the case where $\rho$ is a constant and $Q$ is taken to obey a
power law in $r$, these expressions reduce to those in dFSY.
However, we will not restrict ourselves to this situation.

To eliminate the metric functions $\Phi$ and $\Lambda$ and to get
the basic hydrodynamic equations we consider the conservation laws
$T^{\mu\nu}_{\ \ ;\nu} = 0$.  The fluid part is the same as can be
found in the derivation of the ordinary Oppenheimer-Volkov
equation (see, eg. \cite{Weinberg72}) and is found to be
\begin{equation}
T^{rr}_{\ \ ;r} = e^{-2\Lambda}\left[p' + \Phi'(\rho+p)\right].
\label{divT}
\end{equation}
The electromagnetic part is found to be
\[
(T_{EM}^{rr})_{;r} =
                    -\frac{e^{-2\Lambda}QQ'}{4\pi r^4}.
\]
Setting the sum of these terms to zero and solving for $\Phi'$
yields
 \begin{equation}
  \Phi' = \frac{1}{\rho +
p}\left[\frac{QQ'}{4\pi r^4} - p'\right].
      \label{Phi'}
\end{equation}
Solving (\ref{11}) for $\Phi'$ and inserting the result into
(\ref{Phi'}) yields
\begin{equation}
-\frac{p'}{\rho + p} = -\frac{1}{2r} + e^{2\Lambda}
        \left[4\pi r p + \frac{1}{2r} - \frac{Q^2}{2r^3}\right]
        - \frac{QQ'}{4\pi r^4 (\rho + p)}.
\end{equation}
With the form (\ref{eL}) for $e^{-2\Lambda}$ and some
rearrangement of terms this can be written as
\begin{equation}
p' = \frac{QQ'}{4\pi r^4} - \frac{(\rho + p)}{r^2}
        \left[4\pi r^3p + m(r) + \frac{{\cal F}}{2} -
        \frac{Q^2}{2r}\right]\left(1 - \frac{2m(r)}{r} -
        \frac{{\cal F}}{r}\right)^{-1} \label{OV}.
\end{equation}
This is the Oppenheimer-Volkov (OV) equation for relativistic
charged spheres, which will be the basis for our analysis.

We point out that the Newtonian version of this equation is
\[
p' = \frac{QQ'}{4\pi r^4}  - \frac{\rho m(r)}{r^2},
\]
which shows that in the Newtonian limit the Coulomb repulsion
opposes the gravitational force, as expected, helping to stabilize
charged spheres against collapse. However in the relativistic OV
equation with power-law charge distribution $Q\sim r^k$, ${\cal F}/2
- Q^2/(2r) < 0$.  Thus these terms decrease the effective mass in
the numerator of the second term, tending to ``stabilize." At the
same time charge decreases the denominator, tending to
``destabilize." The final outcome is not entirely clear, which is
why one needs to investigate the OV equation numerically.

%-----------------------------------------------------------------------------
\section{Integration of the Oppenheimer-Volkov Equation}
 \setcounter{equation}{0} \label{sec3}
%-----------------------------------------------------------------------------
The basic strategy is simply to integrate (\ref{OV}) and test for
the stability of solutions.  Before doing this one must specify
the charge distribution and an equation of state. We have no
serious models of any sort for charged ``stars," or even know
whether such objects exist. It seems intuitively reasonable that
due to electrical repulsion the charge distribution should be
weighted toward the surface, but otherwise all choices are
basically {\it ad hoc}. One can imagine several plausible
distributions.  dFSY confined themselves to the simplest case, $Q
\propto r^k$.  We consider two distributions:
\begin{equation}
Q(r) = Q_T \left(\frac{r}{R}\right)^k~e^{cr/R}
     \equiv \sqrt{\beta}~r^k~e^{cr/R},\label{exp}
\end{equation}
where $R$ and $\beta$ are adjustable input parameters that define
the radius of the star and the total charge $Q_T$ respectively,
and
\begin{equation}
Q(r) = \sqrt{\beta} \tanh^k(\frac{r}{cR}),\label{tanh}
\end{equation}
which flattens out for $r \gg cR$. Note that in both these cases
the charge density, obtained by differentiating (\ref{Q}),
diverges at the origin unless $k \ge 3$.  In point of fact,
although we have been able to find solutions for the charge
distribution (\ref{tanh}), the stability analysis is generally
inconclusive since no absolute boundary separation can be found
consistently over all $\beta$ of a given $\gamma$ across the
entire range of parameters considered here. The one exception is
$\gamma=4$ in which case the results are qualitatively similar to
the exponential distribution (\ref{exp}) for all $\beta$, and so
we confine ourselves to that case.

Scarcely less arbitrariness applies to the equation of state (EOS)
than to the charge distribution. Three obvious possibilities
suggest themselves: $\rho = constant$, the adiabatic EOS $p\propto
\rho_{rm}^\gamma$ with arbitrary adiabatic index $\gamma$, and the
ultrarelativistic EOS, $p=\rho/3$. However, despite the fact that
an exact solution with infinite central density $\rho \propto
r^{-2}$ for the last case is known when $Q = 0$ (see \cite
{Weinberg72}), we have found it difficult to find convergent
solutions for this EOS when $Q \ne 0$ when iterating over central
pressure. (This is as opposed to iterating over central density,
which we do for arbitrary adiabatic index, below). Thus we confine
ourselves to the first two choices. We do, however, consider the
$\gamma = 4/3$ adiabatic EOS, which includes relativistic matter
with non-negligible rest mass density (typically $\rho_{rm}/e \sim
10^2 - 10^4$ at the origin for $\gamma = 4/3$, depending on $p_c$
and $\beta$, and substantially lower ratios for larger $\gamma$).
We reiterate that $\rho$ represents the total mass-energy density
of the matter plus gravitational field. Generally, the EOS is
assumed to relate only the rest energy density, $\rho_{rm}$, and
the pressure. As mentioned in \S \ref{sec2}, the total local
energy density is $\rho = \rho_{rm} + e$, where $e$ is the
internal energy density. We consider only polytropes, for which $e
= p/(\gamma-1)$.

Once the charge distribution and EOS are specified we can
integrate the OV equation.  For virtually any case one can
imagine, however, integrating (\ref{OV})  is impossible
analytically and the integration must be carried out numerically.
Even then, except for the case $\rho = constant$, this turns out
to be nontrivial. For $\rho = constant$, the total mass will be
simply $m(r) = (4/3)\pi r^3 \rho$. With the boundary condition
$p(R) = 0$ and a form of $Q$ to make $\cal{F}$ analytically
integrable, this allows convenient integration from the outer
boundary inward. For more reasonable equations of state, however,
one expects $\rho$ to vanish at $r = R$ as well (eg., if $\rho = k
p^{1/\gamma} + p/(\gamma-1)$). In such cases the entire second
term in (\ref{OV}) vanishes at the outer boundary, which causes a
serious problem numerically because $dp/dr|_{r=R} = QQ'/(4\pi r^4)
> 0$, while $p(R-\Delta r) \approx -QQ'|\Delta r|/(4\pi R^4)$.
Thus, unless $QQ' < 0$, an inward integration scheme immediately
predicts a negative pressure unless the integration is initiated
from some arbitrary outer pressure and surface radius values to
compensate for the charge distribution.

For this reason, we choose to integrate outwards from $r = 0$.
This strategy also presents difficulties.  We can, say, choose $R$
as the prespecified ``radius" of the star, but in starting the
integration at the origin, we have no prior knowledge of which
combination of parameters $\rho_c$ (central density), $p_c$
(central pressure), $\gamma$, $\beta$, $k$, $c$, and $Q$ will
actually produce a star with radius $R$.  In other words, if we
define the physical surface radius $R_p$ to be the point at which
the pressure first goes negative, there is no reason that $R_p$
will equal $R$. We have therefore adopted the following procedure:
For a given parameter set $(R,\beta,\gamma,k,c,Q,\rho_c,p_c)$ we
take a first guess at $\rho_c = 1/(R p_c^{1/\gamma})$ with
arbitrary $p_c$, then iterate on $\rho_c$ using a bisection
algorithm until $R_p$ converges to $R$. The bisection method
assumes a locally monotonic correlation between
$\delta\rho_c/\rho_c$ and $\delta R/R$, where the $(n+1)$-th
estimate of the central density is given by $\rho_c^{(n+1)} =
\rho_c^{(n)} + \delta\rho_c^{(n)}$. One chooses
$\delta\rho_c^{(n)}$ by $\rho_c^{(n)}\delta R/R = \sigma
\rho_c^{(n)}(R_p - R)/R$, where $\sigma$ is a coefficient
initially set to unity, but is halved for convergence whenever
$\delta R$ changes sign. We find in most cases we can converge on
$R$ to a relative tolerance of $\delta\rho/\rho \le 0.1\%$.

\begin{figure}[htb]
\vbox{\hfil\scalebox{0.7} {\includegraphics{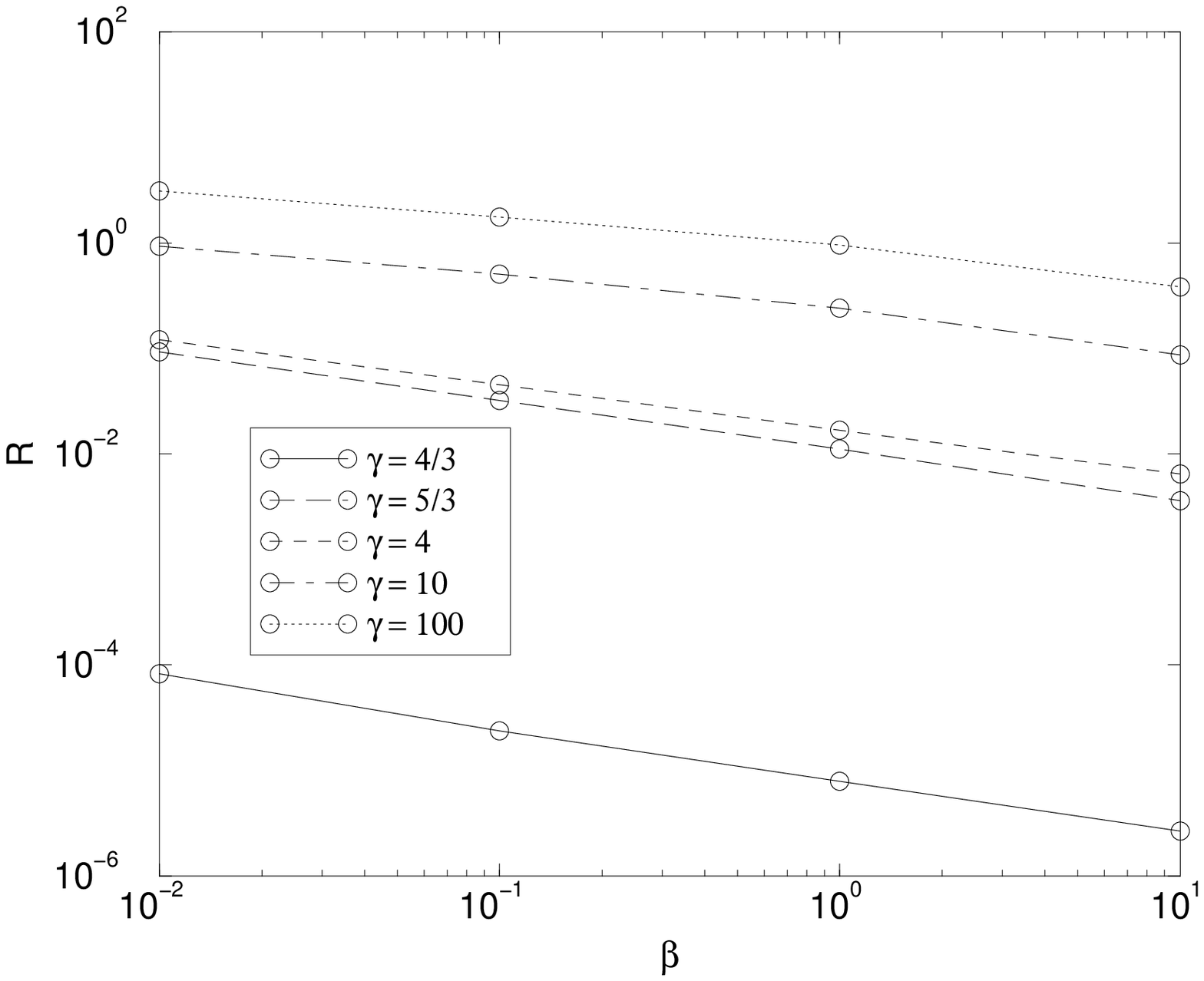}}\hfil}
\caption{ Maximum surface radius as a function of charge parameter
$\beta$ found by converging in the central density to a tolerance
of $\Delta \rho_c/\rho_c = 10^{-3}$ for a charge distribution
$Q=\sqrt{\beta} r^3$ and the different adiabatic indexes
considered in this paper. For fixed $\beta$, the numerical
solutions do not converge to the specified tolerance at radii
above the plotted curves. } \label{fig:rvsbeta}
\end{figure}

The numerical integration is performed using an adaptive fourth
order Runge-Kutta scheme to maintain a constant error tolerance of
$10^{-12}$ over a grid constructed with cell spacing scaled to the
surface radius such that $\Delta r/R \le 5\times10^{-4}$. Further
accuracy is achieved at the outer surface where pressure and
density vary sharply by adaptively refining the mesh spacing to a
minimum level $\Delta r \approx 10^{-12}$ when the pressure is
observed to change sign. This procedure works well if $R$ is
sufficiently small. However, as $R$ is increased one eventually
hits a barrier $R=R_{max}$ beyond which convergence to the $0.1\%$
tolerance is impossible or $(2m(r) + {\cal F})/r$ becomes so large
that the metric function (\ref{eL}) becomes negative. This latter
marks the black-hole limit $R = R_+ = M + \sqrt{M^2 - Q^2}$, where
$M$ is the exterior gravitational mass. The upper bound on
$R_{max}$ generally depends on all input parameters, and is
computed numerically for each case presented in this paper. For
example, considering an adiabatic EOS with $Q=\sqrt{\beta}r^3$,
Figure \ref{fig:rvsbeta} shows the maximum surface radius as a
function of $\beta$ for the different $\gamma$ considered in this
paper. We are unable to find convergent solutions at radii above
each of the curves.

\begin{figure}[htb]
\vbox{\hfil\scalebox{0.7} {\includegraphics{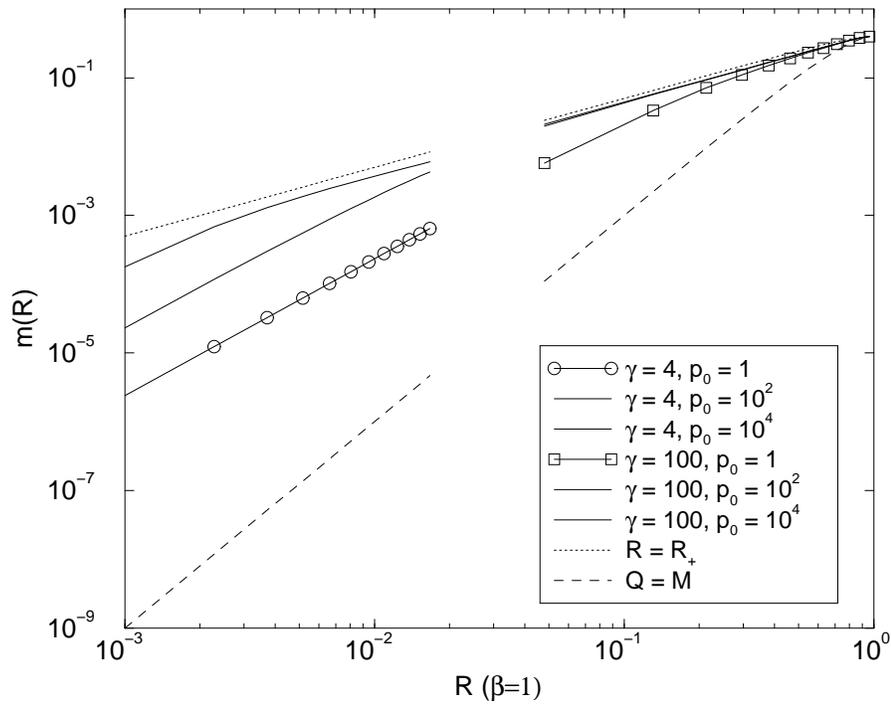}}\hfil}
\caption{ Total mass density $m$ as a function of outer surface
radius $R$ (up to $R=R_{max}$) for fixed charge parameter
$\beta=1$. Each of the solid lines are solutions to the OV
equations with different central pressures (the larger pressures
correspond to greater mass densities). Also plotted are the
densities for the corresponding black hole
($R=R_+$) and extremal charge
($Q=M$) limits. } \label{fig:mvsr-p0}
\end{figure}

Figure \ref{fig:mvsr-p0} plots the total mass $m(R)$ as a function of
surface radius for $\gamma=4,~100$ and fixed $\beta=1$. Here the
surface radius is increased from 5\% of $R_{max}$ to the maximum
computed value $R_{max}$. Each of the solid lines represent
different central pressures $p_c$ with the larger pressures
resulting in greater masses. Also plotted in Figure
\ref{fig:mvsr-p0} are the corresponding black hole and extremal
charge solutions for comparison. The black hole solution for $r =
R$ is found by setting the metric function (\ref{eL}) equal to
zero at that point, which gives $m(R) = (R-{\cal F}(R))/2$. By
contrast, the extremal solution (not necessarily a black hole) is
found by first matching the exterior Reissner-Nordstr\"om metric
to the interior solution (\ref{metric}) at $r=R$:
\begin{equation}
1 - \frac{2M}{R} + \frac{Q^2}{R^2}
 = \frac{1}{R}\int_0^R(1 - 8\pi\rho r^2 -
 \frac{Q^2}{r^2})dr\;\label{match}
\end{equation}
which gives the definition of gravitational mass
\begin{equation}
M = \frac{1}{2}\int_0^R( 8\pi\rho r^2 + \frac{Q^2}{r^2})dr +
\frac{Q^2}{2R}. \label{gravmass}
\end{equation}
Setting $Q = M$ gives the extremal curve for $m(R)= 4\pi\int \rho
r^2 dr$:
\begin{equation}
m(R) = Q - \frac{Q^2}{2R} - \frac{{\cal F}(R)}{2}.\label{excurve}
\end{equation}

\begin{figure}[htb]
\vbox{\hfil\scalebox{0.7} {\includegraphics{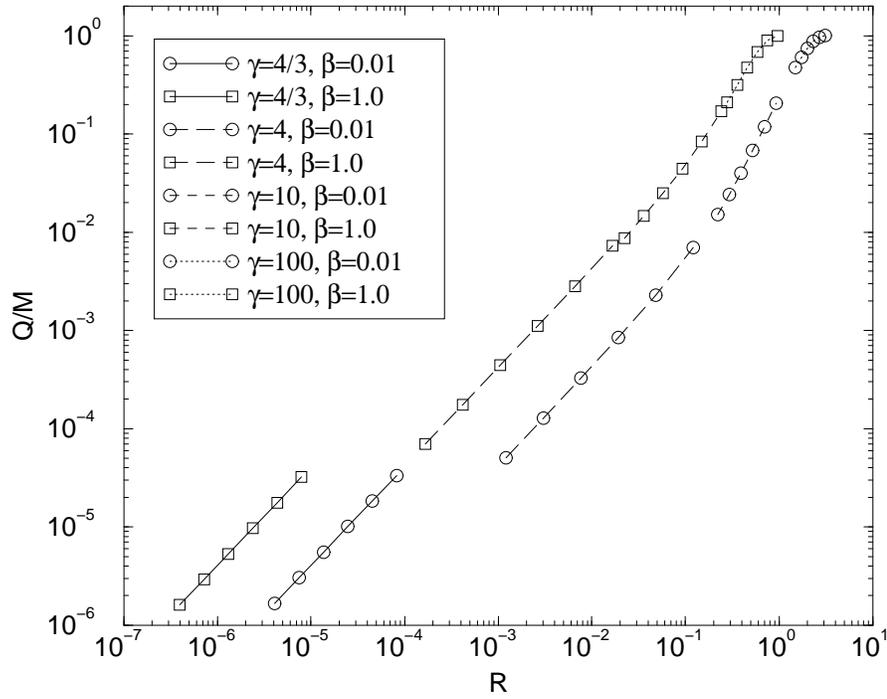}}\hfil}
\caption{ Total charge to mass ratio $Q/M$  is a monotonic
function of surface radius up to the maximum computed radius
$R_{max}$, and peaks at $Q\sim M$ for large $\gamma$. The results
are shown for a central pressure $p_c=1$. } \label{fig:qmvsr}
\end{figure}

Finally we note that the total charge to total mass ratio $Q/M$
generally increases monotonically with increasing $R$ and fixed
$\beta$, as Figure \ref{fig:qmvsr} demonstrates. Since we are
interested in the maximum or extremal limit of $Q/M$, the
subsequent stability plots presented in
section \ref{sec4} are computed for those parameter values
at the upper bounds of $R$ ($= R_{max}$) for which we are able to
find convergent solutions.

As a reminder, we point out that because we do not know beforehand
the exterior mass of the star, the radius $R$ in our graphs is not
given in units of $M$.  $R$ is merely where the pressure drops to
zero expressed in geometric units.

%-----------------------------------------------------------------------------
\section{Stability I: Normal Mode Analysis}
 \setcounter{equation}{0} \label{sec4}
%-----------------------------------------------------------------------------
We provide two stability analyses.  The first, discussed in detail
in this section, is a normal-mode analysis for radial pulsations
of the charged sphere, similar to that performed by dFYS (see also
\cite{Misner73} or \cite{Shapiro83}). The second, discussed in \S
\ref{sec5}, is based on the variation of energy.

We consider radial pulsations of the sphere with the goal of
writing the Sturm-Liouville equation:
\begin{equation}
(F\zeta')' + (H + \omega^2 W)\zeta = 0.
\end{equation}
Here $\zeta$ is a ``renormalized displacement function," and $F,
H$ and $W$ are functions of the zeroeth-order (equilibrium)
solution to the Einstein equations. $\omega$ is the angular
frequency of the assumed sinusoidal time dependence of $\zeta$.
Once these functions are known, we evaluate
\begin{equation}
\omega^2 = \frac{\int_0^R (F\tilde\zeta'^2 - H \tilde\zeta^2) dr}
                {\int_0^R W\tilde\zeta^2 dr} , \label{omega}
\label{omega2}
\end{equation}
and deduce that the solution is stable under radial perturbations
if $\omega^2 > 0$ for approximate trial functions $\tilde{\zeta}$
that satisfy the same boundary conditions as $\zeta$.

To derive $F$, $H$, and $W$ is a tedious undertaking; because we
are now dealing with a time-dependent problem, the metric
functions in (\ref{metric}) must be taken to be $\Phi = \Phi_0(r)
+ \Phi_1(r,t)$ and $\Lambda = \Lambda_0(r) + \Lambda_1(r,t)$,
where we assume the first-order quantities are small.  Once the
perturbation equations are derived, to get the Sturm-Liouville
form, the first-order quantities are eliminated in favor of the
zeroeth-order solutions.  Nevertheless, because our equations
differ slightly from dFYS, we now outline the procedure.  The
development basically follows that of MTW \cite{Misner73}, chapter
26.

We first consider the total energy-momentum tensor. With the
condition $g_{\mu\nu} u^\mu u^\nu = -1$, and raising and lowering
with the full metric, one gets to first order $u_0 = -e^{\Phi_0}
(1+\Phi_1)$ and $u_1 = e^{2\Lambda_0 - \Phi_0} \dot{\xi}$, where
$\xi(r,t)\ll r$ represents the small displacement of a fluid
element initially at radius $r$. For the electromagnetic part of
$T_{\mu\nu}$, we let $Q = Q_0 + Q_1$, with $Q_1 = - Q_0'(r)
\xi(r,t)$ to first order. This is equivalent to assuming that the
charge distribution in the oscillating system is the same as in
the unperturbed system, i.e, that the Lagrangian perturbations
vanish and no electric currents are introduced for the comoving
observer. Then,
\begin{eqnarray}
T_{00} &=& \left(\rho_0 + \frac{Q_0^2}{8\pi r^4}\right)e^{2\Phi_0}
        + e^{2\Phi_0} \left(\rho_1 + \Phi_1 (\frac{Q_0^2}{4\pi r^4}
                                   + 2\rho_0)
        - \frac{Q_0 Q_0'\xi}{4\pi r^4}\right) \nonumber \\
T_{01} &=& -e^{2\Lambda_0}(p_0 + \rho_0) \dot\xi \nonumber\\
T_{11} &=& \left(p_0 - \frac{Q_0^2}{8\pi r^4}\right)e^{2\Lambda_0}
        + e^{2\Lambda_0} \left(p_1 + \Lambda_1(2 p_0 - \frac{Q_0^2}{4\pi r^4})
        + \frac{Q_0 Q_0'\xi}{4\pi r^4}\right). \label{Tpert}
\end{eqnarray}
Note the presence of $T_{01}$, due to the fluid motion.  This is
entirely a first-order quantity.

The left-hand-side of the Einstein equations are most easily
computed with the help of Mathematica or Maple. Then the $(00)$
Einstein equation is found to be
\begin{eqnarray}
&&\frac{e^{2\Phi_0}}{r^2}(1+ 2\Phi_1) +
\frac{2}{r^2}e^{-2\Lambda_0 + 2\Phi_0}\left[-\frac{1}{2} +
r\Lambda_0' + \Lambda_1(1-2r\Lambda_0') -
\Phi_1(1-2r\Lambda_0')+r\Lambda_1'\right]\nonumber \\
  && = 8\pi e^{2\Phi_0} (\rho_0 + \frac{Q_0^2}{8\pi r^4})
     + 8\pi e^{2\Phi_0} \left(\rho_1 + 2\Phi_1(\rho_0
          + \frac{Q_0^2}{8\pi r^4})
          - \frac{2Q_0Q_0'\xi}{8\pi r^4}\right)
\end{eqnarray}
Equating orders on each side leads to the $0^{th}$-order equation
(\ref{00}), which minus the charge term is also MTW Eq. (26.1b).
The first-order equation, after some simplification, comes out to
be
\begin{equation}
\frac{2e^{-2\Lambda_0}}{r^2}\left[\Lambda_1(1-2r\Lambda_0')
    + r\Lambda_1'\right] = 8\pi\rho_1 - \frac{2Q_0Q_0'\xi}{r^4}.
\end{equation}

With  $T_{01}$ given by (\ref{Tpert}), the (01) Einstein equation
is then nontrivial and gives
\begin{equation}
\Lambda_1 = - 4\pi r e^{2\Lambda_0} (p_0 + \rho_0)\xi
          = -(\Lambda_0' + \Phi_0')\xi, \label{L1}
\end{equation}
where the second equality follows from the zeroeth-order equations
(\ref{00}) and (\ref{11}).

We will shortly need an expression for the perturbed pressure
$p_1$. One can get this from the law of baryon conservation
$d(\Delta n)/d\tau = -n {\bf(\nabla \cdot u)} =
(-g)^{-1/2}(\sqrt{-g} u ^\alpha),_\alpha$. Taking $\alpha = t, r$
yields
\begin{equation}
\Delta n = -n_0\left[r^{-2} e^{-\Lambda_0}(r^2 e^{\Lambda_0} \xi)'
+\Lambda_1\right],\label{dn}
\end{equation}
where $\Delta n$ represents the Lagrangian perturbation in $n$.
Assuming adiabatic fluctuations such that $\Delta p/\Delta n =
\gamma p/n$ then gives
\begin{equation}
p_1 = -\gamma p_0 \left[r^{-2} e^{-\Lambda_0}(r^2 e^{\Lambda_0}
\xi)' +\Lambda_1\right] - \xi p_0', \label{p1}
\end{equation}
which is MTW Eq. (26.9).

The first-order (11) Einstein equation is found to be
\begin{equation}
\Phi_1' = 4\pi r e^{2\Lambda_0} \left(p_1 + 2\Lambda_1(p_0 -
\frac{Q_0^2}{8\pi r^4})\right)
            + \frac{\Lambda_1 e^{2\Lambda_0}}{r}
            + e^{2\Lambda_0} \frac{Q_0 Q_0'\xi}{r^3}.
\end{equation}
With (\ref{L1}) and (\ref{p1}), $\Lambda_1$ and $p_1$ can be eliminated to get
\begin{eqnarray}
 \Phi_1' &=& -4\pi r e^{2\Lambda_0}
        \left[\gamma p_0 r^{-2} e^{-\Lambda_0} (r^2 e^{\Lambda_0} \xi)'\right]
        \nonumber \\
       && -4\pi r e^{2\Lambda_0}\xi
        \left[p_0' + (\Lambda_0' + \Phi_0')
                      \left(2p_0 -\gamma p_0 - \frac{Q_0^2}{4\pi r^4}
                                             + \frac{1}{4\pi r^2}\right)
                    - \frac{Q_0 Q_0'}{4\pi r^4} \right]
\end{eqnarray}

By considering the parallel component of the conservation equation
$u_\mu T^{\mu\nu}_{\ \ ;\nu} = 0$, using (\ref{dn}) and assuming
the Lagrangian perturbation $\Delta Q$ vanishes, one can show
\begin{equation}
\rho_1 = -(p_0 + \rho_0)
         \left[r^{-2}e^{-\Lambda_0}(r^2e^{\Lambda_0}\xi)' + \Lambda_1\right]
         -\xi \rho_0',
\end{equation}
which is MTW Eq. (26.11).

The transverse component of the conservation equation, $h^\mu_\nu
T^{\sigma}_{\ \mu;\sigma} = 0$, where the projection tensor
$h^\mu_\nu = \delta^\mu_\nu + u^\mu u_\nu$ eventually yields
\begin{eqnarray}
e^{2\Lambda_0 - 2\Phi_0}(\rho_0 + p_0) \ddot{\xi} &=&
      -\Phi_0'(\rho_1 + p_1) - \Phi_1'(\rho_0 + p_0) - p_1' \nonumber \\
   && -\frac{1}{4\pi r^4}\left(Q_0'^2\xi + Q_0Q_0''\xi + Q_0Q_0'\xi'\right)
\end{eqnarray}
We note that the $Q_0''$ term does not appear in dFSY.

With these expressions we are finally able to write down the
Sturm-Liouville equation.  We define the renormalized displacement
function $\zeta = r^2 e^{-\Phi_0} \xi$ and assume $\zeta =
\zeta(r) e^{-i\omega t}$.  Then
\begin{equation}
\begin{array}{lll}
 F &=& \gamma p_0 r^{-2} e^{\Lambda_0 + 3\Phi_0},
                            \nonumber \vspace{1 mm}\\
 W &=& r^{-2} (\rho_0 + p_0) e^{3\Lambda_0 + \Phi_0},\nonumber\vspace{1 mm} \\
H &=& r^{-2} e^{\Lambda_0 + 3\Phi_0}
      \left[ 4\pi r(\rho_0 + p_0) e^{2\Lambda_0}
             \left(-p_0' + \frac{Q_0 Q_0'}{4\pi r^4} + \frac{1}{r}(\rho_0 + p_0)\right) \right.
             \nonumber \vspace{2mm} \\
&&\left.+ 2p_0'\Phi_0' + p_0'' + \Phi_0'\rho_0'
  -\frac{2p_0'}{r}
     -  \frac{1}{4\pi r^4}(Q_0'^2 + Q_0 Q_0'' +
                                Q_0Q_0'(\Phi_0' - \frac{2}{r}))
      \right].
\end{array}
\end{equation}
 We have not attempted to reconcile this expression for
$H$ with the one in dFSY.

Physically reasonable solutions require $\zeta/r^3$ finite or zero
as $r \to 0$ and $\gamma p_0r^{-2}e^{\Phi_0}\zeta' \to 0$ as $r
\to R$.  Any trial function $\zeta$ that satisfies these
conditions and that extremizes (\ref{omega}) is acceptable.
Following Chandrasekhar and dFSY, we choose $\zeta \propto r^3$.

\begin{figure}[htb]
\vbox{\hfil\scalebox{0.7}
{\includegraphics{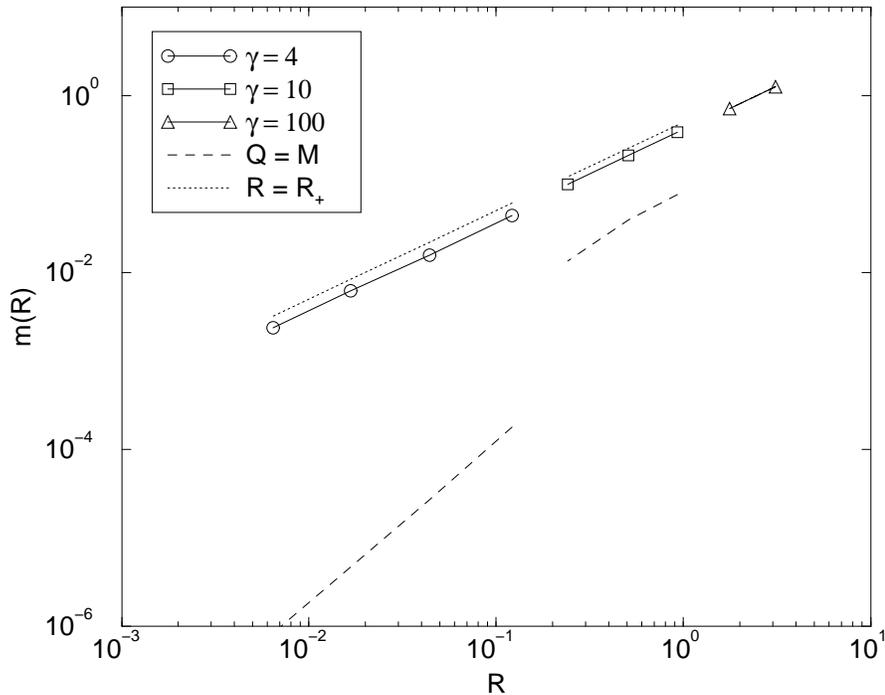}}\hfil} \caption{ Total
mass function corresponding to the stability limit as a function
of the maximum computed surface radius for $\gamma=4,~10,~100$.
The solid lines are found by increasing the central pressure from
unity and plotting the first points which become unstable. The
four points in the $\gamma=4$ curve represent for this
distribution $Q=\sqrt{\beta}r^3$, $\beta=0.01,~0.1,~1.0,~10$
(right to left).  The three points for $\gamma=10$ represent
$\beta=0.01,~0.1,~1.0$; the two points for $\gamma=100$ represent
$\beta=0.01,~0.1$. The displayed data are chosen to prevent
overlap of the different $\gamma$ curves, but we note that the
general conclusion remains the same for all cases and choices of
$\beta$ investigated: the limit of stability lies between the
$Q=M$ curves (dashed lines) and the black hole $R=R_+$ limit (dotted
lines), even for $\gamma=4/3$ and 5/3 which are not shown here for
clarity. Thus instability sets in before extremality is reached.
} \label{fig:stab-max-adiabatic}
\end{figure}

Figure \ref{fig:stab-max-adiabatic} shows the stability curves for
the adiabatic EOS with $Q=\sqrt{\beta}r^3$. The results are
generated by choosing an initial central pressure (equal to unity)
and iteratively solving the OV equation to convergence in the
central density for fixed $\beta$ and surface radius $R=R_{max}$.
The central pressure is then monotonically increased and the OV
equation repeatedly solved until the angular frequency defined in
equation (\ref{omega2}) becomes negative. The data points
connected by solid lines are the solutions at which the frequency
first becomes negative. Also shown are the black hole limit $R =
R_+$ and the $Q=M$ limit, generated from Eq. (\ref{excurve}). Note
that the stability limit falls {\it between} the $Q = M$ curve and
the black-hole curve.  As can be seen from Figure \ref{fig:qmvsr},
$Q/M$ varies over a broad range at $R=R_{max}$, but is constrained
by $Q/M < 1$ in the limit of large $\gamma$. Thus instability sets
in {\it before} extremality is reached since $Q/M$ decreases as
the $R=R_+$ curve is approached (say by increasing the central
pressure), even in the limit $\gamma \to \infty$ where all the
curves come together.

\begin{figure}[htb]
\vbox{\hfil\scalebox{0.7}
{\includegraphics{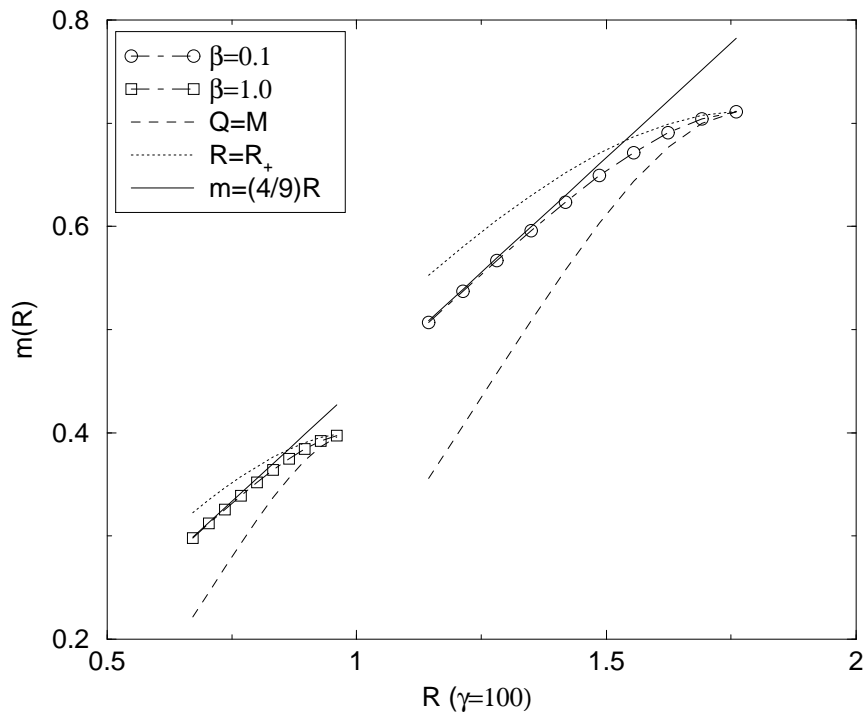}}\hfil} \caption{ As Figure
\protect{\ref{fig:stab-max-adiabatic}}, except here we show
closeups of the extremal, black hole, and stability curves for the
more interesting high adiabatic index ($\gamma=100$) case over a
range of surface radii leading up to the maximum $R=R_{max}$. Also
plotted is the solid line $m(R) = 4R/9$. The usual stability limit
for the interior Schwarzschild solution is actually given by $M =
4R/9$, where M is the gravitational mass.  The two expressions are
equivalent only in the limit $Q = 0$; 
see also Figure \protect{\ref{fig:stab-adiabaticM}}. In all
cases, the stability curves separate the extremal charge and black
hole states. } \label{fig:stab-adiabatic}
\end{figure}

\begin{figure}[htb]
\vbox{\hfil\scalebox{0.7}
{\includegraphics{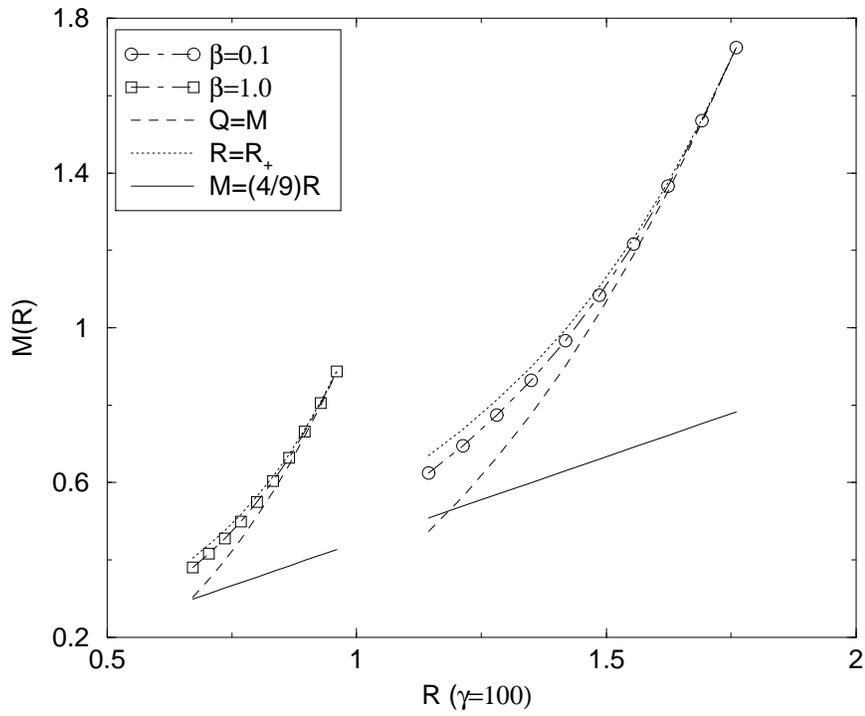}}\hfil} \caption{ As Figure
\protect{\ref{fig:stab-adiabatic}}, except here we plot solutions
for the gravitational mass $M$ as a function of surface radius
$R$. This time the solid lines show the stability limit $M(R) =
4R/9$ for the interior Schwarzschild solution. Notice there is no
absolute upper bound on stability other than the black hole limit
$R \to M$ as $R\to R_{max}$ and $Q\to M$. }
\label{fig:stab-adiabaticM}
\end{figure}

Although for reasons of clarity we have not shown the $\gamma=4/3$
and 5/3 solutions in Figure \ref{fig:stab-max-adiabatic}, they are
qualitatively similar to the other cases, except that the
extremal, stability and black hole curves are more widely
separated. A more detailed plot of the $\gamma=100$ case is shown
in Figure \ref{fig:stab-adiabatic} for two different values of
$\beta$, and varying the surface radius up to $R=R_{max}$. Also
shown in Figure \ref{fig:stab-adiabatic} with solid lines are the
corresponding limits for the interior Schwarzschild solution when
$Q=0$, $m(R) = 4R/9$. This absolute limit is a good match to the
numerically computed results even for relatively large ratios of
$Q/M$ and radii $R\sim R_{max}$. However, a more useful extension
of this limit should employ the gravitational mass $M$ deriving
this constraint. Figure \ref{fig:stab-adiabaticM} plots the same
result as \ref{fig:stab-adiabatic}, but for the gravitational mass
$M(R)$. The solid lines are now the limits for the exterior
Schwarzschild solution $M(R) = 4R/9$. Notice that the upper bound
of stability with this definition is always below, but
approaches the black hole limit
$R\to R_+ \to M$ as $Q\to M$ and $R\to R_{max}$.
This result is consistent with that
of \cite{yunqiang99} who find no non-singular static solutions
with $R\to R_+$ for $Q<M$.

\begin{figure}[htb]
\vbox{\hfil\scalebox{0.7}
{\includegraphics{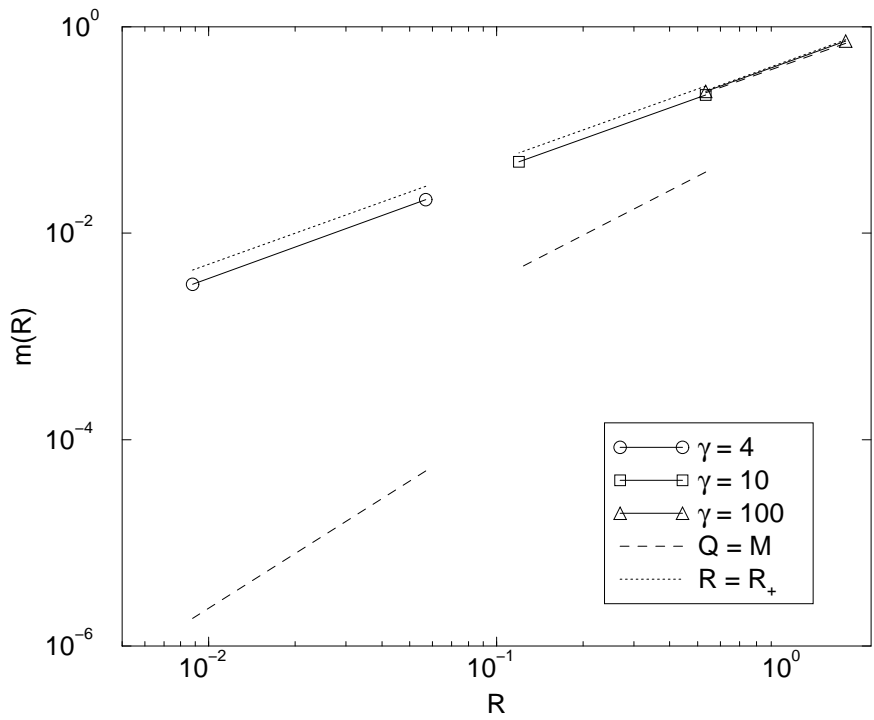}}\hfil} \caption{ As Figure
\protect{\ref{fig:stab-max-adiabatic}}, but for a charge
distribution of $Q=\sqrt{\beta}r^3e^{r/R}$. The symbols represent
$\beta= 0.01$ (right-most points) and 1.0 (left-most points). }
\label{fig:stab-max-expq}
\end{figure}

\begin{figure}[htb]
\vbox{\hfil\scalebox{0.7}
{\includegraphics{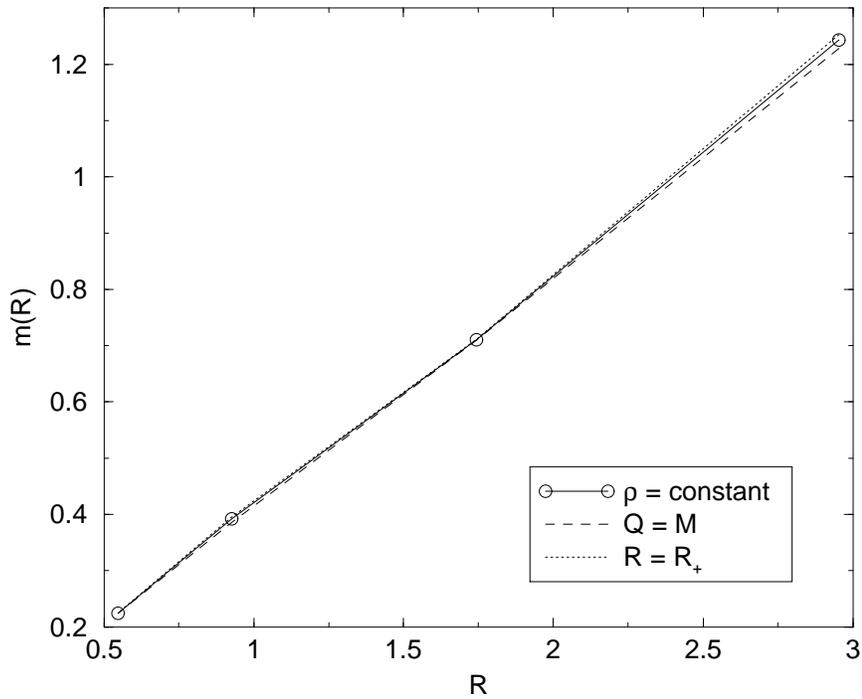}}\hfil} \caption{ As
Figure \protect{\ref{fig:stab-max-adiabatic}}, but for a constant
density distribution. The symbols represent $\beta=
0.01,~0.1,~1.0,~10$ (right to left). Although difficult to see, an
unstable solution (sold line) always lies between the extremal and
black hole curves for all surface radii $R\le R_{max}$ we have
studied. This figure is shown for $R = R_{max}$ as are most of the
other stability plots. } \label{fig:stab-max-constantrho}
\end{figure}

The overall behavior is repeated in Figures
\ref{fig:stab-max-expq} and \ref{fig:stab-max-constantrho}, which
show similar graphs for two additional cases: the adiabatic EOS
with $Q=\sqrt{\beta}r^3e^{r/R}$, and a constant density profile
respectively. Once again, unstable solutions separate the $Q=M$
and black hole states. This is our main result: that the onset of
instability always occurs before extremality is reached.

%-----------------------------------------------------------------------------
\section{Stability II: Energy Analysis}
\begin{figure}[htb]
\vbox{\hfil\scalebox{0.7} {\includegraphics{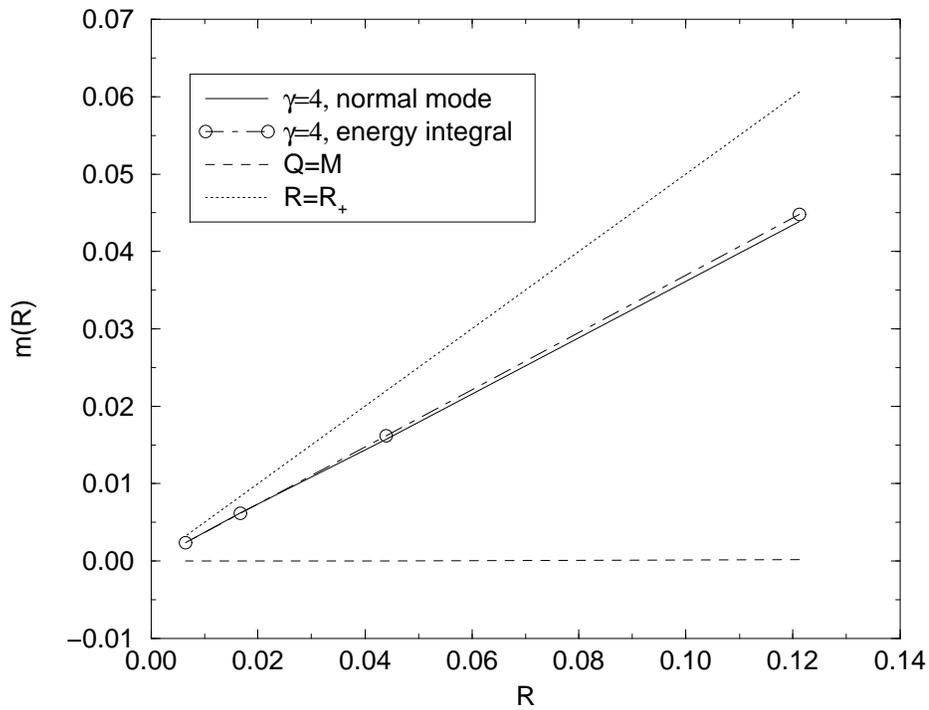}}\hfil}
\caption{Comparison of the energy variation and normal mode
methods in evaluating the stability of solutions for the same
$\gamma=4$ case as Figure \protect{\ref{fig:stab-max-adiabatic}},
but displayed on a linear-linear scale to distinguish the
stability lines. Over the parameter range that both methods can be
applied to reliably (mostly restricted by the energy method which
is less robust from a numerical point of view), the results agree
remarkably well. } \label{fig:energyintegral}
\end{figure}

 \setcounter{equation}{0} \label{sec5}
%-----------------------------------------------------------------------------
Admittedly, the above stability analysis is not exceptionally
transparent.  Somewhat more insight into the stability of
relativistic charged spheres can be gained by considering an
energy analysis.  Following \cite{Shapiro83} and neglecting the
electrostatic energy which is invariant with respect to changes in
density, one can write the internal (thermal plus gravitational)
energy of a spherical gas cloud as
\begin{equation}
E = \int_0^R\left[\rho \left(1-\frac{2m(r)}{r} - \frac{{\cal
F}(r)}{r}\right)^{1/2} - \rho_{rm} \right] d{\cal V},
\label{energyintegral}
\end{equation}
where the invariant volume element d${\cal V} = 4\pi(1-2m(r)/r -
{\cal F}(r)/r)^{-1/2}r^2 dr$. An equilibrium configuration is
determined by the condition $\partial E/\partial\rho_{rmc} = 0$,
where $\rho_{rmc}$ is the central rest-mass density, and the
transition from a stable to unstable configuration is determined
by the condition $\partial^2E/\partial\rho_{rmc}^2 = 0$ at the
point where $\partial E/\partial\rho_{rmc} = 0$. We find the
transition numerically  by first iteratively solving the OV
equation for a given central pressure $p_c$ until convergence is
reached in $\rho_{c}$.  The energy integral (\ref{energyintegral})
is then evaluated and tabulated as a function of total mass and
central rest-mass density. The inflection point is then easily
computed as a function of $M$ by discretizing the tabulated data
with respect to $\rho_{rmc}$. Figure \ref{fig:energyintegral}
shows a comparison of this method of evaluating stability with the
normal mode analysis of \S \ref{sec4} for the $\gamma=4$ case of
Figure \ref{fig:stab-max-adiabatic}, but on a linear-linear scale.
The results agree remakably well. However, we note that this
energy approach is much more sensitive to numerical resolution and
accuracy than the normal mode method. In fact, this method proves
generally inconclusive for extreme (both small and large) values
of $\gamma$. Hence we rely exclusively to the more robust normal
mode method of \S \ref{sec4} to assess stability.

Eq. (\ref{energyintegral}) is already suggestive in that one can
immediately see that the function $\cal{F}$ enters with the same
sign as $m(r)$.  In other words, charge effectively increases the
magnitude of the gravitational potential energy, against the
internal energy included in $\rho$, which should tend to
destabilize the charged sphere.  This can be confirmed by a
perturbation analysis. Let $\rho = \rho_{rm}(1 + u)$, where now
$u$ is the internal energy per unit mass, and assume that both $u
\ll 1$ and $m(r)/r \ll 1$.  Note that ${\cal F}/r \sim Q^2/r^2 =
(Q^2/m^2)(m^2/r^2)$, so this is at least a second order effect,
${\cal F}/r \sim m^2/r^2$ if $Q\sim m$, and even higher order if
$Q\ll m$. Expanding (\ref{energyintegral}) to lowest order in
$\cal{F}$, gives
\begin{equation}
E = \int_0^R\rho_{rm}\left[u -\frac{m}{r} - \frac{um}{r}
    -\frac{m^2}{2r^2} - \frac{{\cal F}}{2r}\right]d{\cal V}.
\end{equation}

The first four terms are those found in computing the standard
relativistic corrections to a Newtonian star (see
\cite{Shapiro83}, chap. 6). The last term, which we denote as $I_Q
\equiv (1/2)\int\rho_{rm}({\cal F}/r) d{\cal V}$ is the correction
due to charge. Note that $I_Q \ge 0$ always. For a $\gamma = 4/3$
polytrope one finds with the correction due to charge
\begin{equation}
E = (AM - BM^{5/3})\rho_{rmc}^{1/3} + CM\rho_{rmc}^{-1/3}
        - DM^{7/3}\rho_{rmc}^{2/3} - I_Q,
\end{equation}
where $A,B,C,D$ are positive numbers that depend on fundamental
constants as well as on results of the numerical integration (see
\cite{Shapiro83}, Eq. (6.10.23)).

The condition $\partial E/\partial\rho_{rmc} = 0$ gives
\begin{equation}
\frac{1}{3}(AM - BM^{5/3})\rho_{rmc}^{-2/3}
     -\frac{1}{3}CM\rho_{rmc}^{-4/3}
        -\frac{2}{3} DM^{7/3}\rho_{rmc}^{-1/3} - I_Q' = 0.
\label{energypert}
\end{equation}
Here, $I_Q' \equiv \partial I_Q/\partial \rho_{rmc} \approx
(1/2)\int(a(r){\cal F}/r) d{\cal V}$, where $a(r)$ represents the
normalized density profile $\rho_{rm} = \rho_{rmc} a(r)$ with
$a(r=0)=1$. We have also neglected higher order contributions from
the invariant volume element $d{\cal V}$. For real stars, equating
the first term in equation (\ref{energypert}) to zero leads to the
Chandrasekhar limit. The remaining terms are small corrections.
Solving for the first term in the above equation and using the
result in the condition $\partial^2E/\partial\rho_{rmc}^2 = 0$
yields
\begin{equation}
\frac{2}{9}CM\rho_{rmc}^{-7/3} -
\frac{2}{9}DM^{7/3}\rho_{rmc}^{-4/3} -
\frac{2}{3}\rho_{rmc}^{-1}I_Q' = 0.
\end{equation}
To leading order $M \approx (A/B)^{3/2}$, so the onset of
instability occurs at
\begin{equation}
C(\frac{A}{B})^{3/2} - D(\frac{A}{B})^{7/2}\rho_{crit}
    - 3\rho_{crit}^{4/3}I_Q' = 0.
\end{equation}
When $I_Q = 0$, this gives the standard result $\rho_{crit} =
(C/D)(B/A)^2$. However, since $I_Q' \ge 0$, the presence of the
last term clearly lowers $\rho_{crit}$, decreasing the threshold
of instability by making the star unstable at lower densities and
masses $m(r)$ at fixed $R$. This behavior is observed in Figure
\ref{fig:stab-adiabatic} where as $Q$ and $Q/M$ are increased the
stability curve falls further below the $R = (9/4) m(R)$
Schwarzschild stability limit. But, since we are increasing
charge, we are increasing the gravitational mass $M$ over
$m(r)$ (see \ref{gravmass}), and so, paradoxically, when measured
in $M$, the stability limit increases, as seen in Figure
\ref{fig:stab-adiabaticM}.

%-----------------------------------------------------------------------------
\section{Conclusions}
 \setcounter{equation}{0} \label{sec6}
%-----------------------------------------------------------------------------
We have seen that, contrary to Newtonian intuition, in general
relativity charge does not tend to stabilize or counterbalance the
interior energy density of fluid spheres. To be sure, our
numerical results indicate that for a fixed radius, instability
sets in at a smaller mass $m(r)$ than in the fluid only case.
Nevertheless, charge does increase the gravitational mass $M$, and
the instability threshold for relativistic charged spheres monotonically
approaches the black hole horizon in the extremal limit. Most important,
though, for all cases studied the onset of instability of
relativistic charged spheres takes place before the attainment of
extremality. At that point, gravitational collapse into black
holes takes place, but these are not extremal holes. All these
results are certainly consistent with the currently accepted form
of the third law of black hole dynamics, which forbids attainment
of the extremal state by finite time processes involving accretion
of positive energy \cite{Israel86} (see also Appendix).

One might legitimately doubt whether studies such as this one have
any bearing on reality given that astrophysical objects tend not
to be charged. For precisely this reason it has long been doubted
that Reissner-Nordstr\"om solutions can represent astrophysical
objects. On the other hand, to the extent that charged solutions
shed light on the third law, they have been of great interest.
Regardless, however, of whether one chooses to interpret charged
spheres as stars or as solitons or as models for elementary
particles, our results make it difficult to see how extremal
objects could arise from any remotely realistic collapse scenario.
A general theorem ruling out stable extremal black holes
altogether would be desirable; it would effectively be an
alternate formulation of the third law, as well as  a strong
version of the cosmic censorship hypothesis (or theorem, in that
case); not only are stable naked singularities
excluded, but stable extremal black holes as well.
\\

\noindent{\bf Acknowledgements} We thank Werner Israel for an
exchange that instigated the appendix. This work was performed
under the auspices of the U.S. Department of Energy by Lawrence
Livermore National Laboratory under Contract W-7405-Eng-48.

%-----------------------------------------------------------------------------
\section*{Appendix: A note on Hawking radiation and the third law}
\setcounter{equation}{0}
\renewcommand{\theequation}{A.\arabic{equation}}
%-----------------------------------------------------------------------------
Israel's proof of the third law \cite{Israel86}, that the surface
gravity of a black hole can never be forced to zero in a
finite-time, continuous process, assumed accreting charged shells
and that the energy flux crossing the outer apparent horizon of
the black hole was positive. Therefore the proof has nothing to
say about Hawking radiation, because in this case a negative
energy flux crosses the apparent horizon, violating the
assumptions of the theorem. Moreover, at first glance, it might
seem that Hawking radiation would drive a black hole toward
extremality: If a black hole is emitting neutral particles, for
example, the charge would remain constant while the mass
decreased, forcing $Q/M \to 1$.  It is not terribly difficult to
see that this does not happen.

The power radiated by the black hole is
\begin{equation}
P = \frac{dM}{dt} \sim R^2T^4 \sim R^2\kappa^4 ,\label{power}
\end{equation}
where $\kappa$ is the surface gravity.  For a Schwarzschild black
hole, $R = 2M$ and $\kappa = 1/4M$.  This immediately gives the
well known result for the evaporation time $\tau \sim M^3$.

For a Reissner-Nordstr\"om black hole, however, $R = R_+ = M +
\sqrt{M^2 - Q^2}$ and the surface gravity is given by
\begin{equation}
\kappa = \frac{(M^2 - Q^2)^{1/2}}{2M[M+(M^2-Q^2)^{1/2}]-Q^2}
\label{kappa}
\end{equation}
(see Wald \cite{Wald84} Eq. (12.5.4)).  Of course, as $Q \to M$,
$\kappa$ approaches zero, and therefore the flux, which goes as
the fourth power of $\kappa$ is rapidly decreasing.  This suggests
that an infinite amount of time is required to reach $Q = M$. The
statement can be made rigorous by integrating (\ref{power}) for
Reissner-Nordstr\"om black holes.  With $R = R_+$ and the above
expression for $\kappa$ we have
\begin{equation}
\tau =  \int {\frac {(2\, M\,(M + \sqrt{M^{2} - Q^{2}}) -
Q^{2})^{4}}{(M + \sqrt{M^{2} - Q ^{2}})^{2}\,(M^{2} - Q^{2})^{2}}}
\,dM, \label{evaptime}
\end{equation}
where the limits should be from $M$ to $Q$, assuming $M > Q$
initially.  With $Q = constant$, the expression can be integrated
analytically.  The result for the indefinite integral is
\begin{eqnarray}
\tau(M) &=& \frac{32 M^3}{3} + 16MQ^2 + \frac{MQ^4}{2(Q^2-M^2)}
\nonumber\\ && + \frac{2(16M^4 + 16M^2Q^2 -
41Q^4)}{3\sqrt{M^2-Q^2}}
   - \frac{35Q^3\tanh^{-1}(M/Q)}{2} .
\end{eqnarray}
Assuming, for example, that $M_{initial} \gg Q$ and $M_{final} =
Q(1+\epsilon)$ with $\epsilon < 1$, then
\begin{equation}
\tau \sim \frac{64 M^3}{3} + \frac{Q^3}{4\epsilon} +
O(\epsilon^{-1/2}) ,
\end{equation}
which clearly diverges in the limit $Q \to M$ (equivalently
$\epsilon \to 0$).

Although we have done this only for $Q = constant$, it may hold
true generally. If the Hawking temperature $T$ is initially below
the rest mass of the electron, the black hole will radiate only
neutral particles and so $Q$ will remain constant while $M$
decreases. Plotting (\ref{kappa}) shows that $\kappa$ and hence
$T$ will actually {\it increase} with decreasing mass until $Q =
(\sqrt{3}/2)M$, at which point $\kappa = 2/(9M)$ then
monotonically decrease to zero as $\kappa\sim \sqrt{2\epsilon}/Q$
in the limit $M=Q(1+\epsilon)$. Assuming the maximum temperature
is always less than the mass of the electron, then $Q$ can be
expected to remain constant, and evaporation cannot proceed in
finite time, as discussed above.

If at any time $T$ becomes greater than $m_e$, the hole will
evaporate charge. But for any charged elementary particle $q \gg
m$ due to negligible size of the gravitational interaction.
Assuming that for temperatures above the rest mass, all species of
particles are radiated with roughly equal probability
(equipartition), under evaporation $Q$ will become negligible
compared to $M$ and extremality will never be approached.

We emphasize that, as Israel points out, the entire discussion
neglects any superradiant component which is nonthermal, but it is
not clear whether this would alter the conclusion. We have also
neglected other processes, such as discharge from pair production
mentioned by Page \cite{Page76}, which in this case would likely
only strengthen the conclusion. In any case, certainly for $Q =
constant$, Hawking radiation cannot violate the third law of black
hole dynamics and perhaps not in general. This is consistent with
the early work of Page and with the point of view that extremal
black holes do not exist.

\end{document}